\documentclass[11pt,letterpaper]{article}
\usepackage[utf8]{inputenc}
\usepackage[T1]{fontenc}
\usepackage[margin=0.75in]{geometry}
\usepackage{mathptmx}
\usepackage{amsmath,amssymb,graphicx,needspace,titlesec}
\usepackage{xcolor}
\usepackage[colorlinks=true,urlcolor=blue,linkcolor=blue,citecolor=blue]{hyperref}
\hypersetup{pdftitle={Systematic study of the performance of small robots on controlled laboratory substrates},pdfauthor={Chen Li; Aaron M. Hoover; Paul Birkmeyer; Paul B. Umbanhowar; Ronald S. Fearing; Daniel I. Goldman}}
\titleformat{\section}{\fontsize{13}{16}\selectfont\bfseries}{}{0pt}{}
\titlespacing*{\section}{0pt}{14pt}{7pt}
\renewenvironment{abstract}{\section*{Abstract}}{\par}
\let\originalthebibliography\thebibliography
\let\endoriginalthebibliography\endthebibliography
\renewenvironment{thebibliography}[1]{\originalthebibliography{#1}\fontsize{9.3}{11.625}\selectfont\setlength{\itemsep}{3pt}\setlength{\parskip}{0pt}}{\endoriginalthebibliography}
\makeatletter
\renewcommand{\@biblabel}[1]{#1.}
\makeatother

\titleformat{\subsection}{\fontsize{11.5}{14}\selectfont\bfseries}{}{0pt}{}
\titlespacing*{\subsection}{0pt}{10pt}{5pt}
\makeatletter
\long\def\@makecaption#1#2{\vskip\abovecaptionskip{\fontsize{9.5}{12.35}\selectfont\noindent\textbf{Figure \thefigure.} #2\par}\vskip\belowcaptionskip}
\makeatother

\newsavebox{\manuscriptimage}
\newcommand{\manuscriptfigure}[4]{%
\par\begingroup
\sbox{\manuscriptimage}{\includegraphics[#2]{#3}}%
\Needspace{\dimexpr\ht\manuscriptimage+\dp\manuscriptimage+45pt\relax}%
\noindent\makebox[\linewidth][c]{\hypertarget{fig#1}{}\usebox{\manuscriptimage}}\par
\vspace{3pt}\refstepcounter{figure}\label{figure#1}%
{\fontsize{9.5}{12.35}\selectfont\noindent\textbf{Figure \thefigure.} #4\par}%
\endgroup\vspace{8pt}}
\begin{document}
\fontsize{11.2}{15.23}\selectfont
{\centering\fontsize{18}{21.6}\selectfont\bfseries Systematic study of the performance of small robots on controlled laboratory substrates\par}
\vspace{8pt}
{\centering\fontsize{11}{15.4}\selectfont Chen Li$^{a}$, Aaron M. Hoover$^{b}$, Paul Birkmeyer$^{b}$, Paul B. Umbanhowar$^{c}$,\\ Ronald S. Fearing$^{b}$, Daniel I. Goldman$^{a}$\par}
{\centering\fontsize{9.3}{12.1}\selectfont
$^{a}$School of Physics, Georgia Institute of Technology, Atlanta, GA 30332, USA\\
$^{b}$Department of Electrical Engineering and Computer Sciences, University of California, Berkeley, Berkeley, CA 94720, USA\\
$^{c}$Department of Mechanical Engineering, Northwestern University, Evanston, IL 60208, USA\par}
\vspace{4pt}
{\fontsize{9.5}{12.825}\selectfont
Published in Proceedings of SPIE 7679, 76790Z (2010).\\
DOI: \href{https://doi.org/10.1117/12.851047}{\underline{10.1117/12.851047}}\par}

\begin{abstract}
The design of robots able to locomote effectively over a diversity of terrain requires detailed ground interaction models; unfortunately such models are lacking due to the complicated response of real world substrates which can yield and flow in response to loading. To advance our understanding of the relevant modeling and design issues, we conduct a comparative study of the performance of DASH and RoACH, two small, biologically inspired, six legged, lightweight ($\sim10$~cm, $\sim20$~g) robots fabricated using the smart composite microstructure (SCM) process.  We systematically examine performance of both robots on rigid and flowing substrates.  Varying both ground properties and limb stride frequency, we investigate average speed, mean mechanical power and cost of transport, and stability. We find that robot performance and stability is sensitive to the physics of ground interaction: on hard ground kinetic energy must be managed to prevent yaw, pitch, and roll instability to maintain high performance, while on sand the fluidizing interaction leads to increased cost of transport and lower running speeds. We also observe that the characteristic limb morphology and kinematics of each robot result in distinct differences in their abilities to traverse different terrains. Our systematic studies are the first step toward developing models of interaction of limbs with complex terrain as well as developing improved limb morphologies and control strategies.
\end{abstract}


\section{Introduction}
\label{sec:intro}  

In nature, animals display high mobility on terrains ranging from hard surfaces like rocks to deformable substrates like sand and rubble~\underline{\cite{stebbins1972amphibians}}, snow~\underline{\cite{parker1984energy}}, or even the surface of water~\underline{\cite{glasheen1996hydrodynamic}}. In contrast, while man made wheeled and treaded vehicles demonstrate excellent mobility and low energy cost on rigid ground like paved roads, they often suffer total performance loss in deformable terrains like sand\underline{\cite{kumagai04}}. The main challenge of moving on deformable ground is that the locomotor can experience both solid- and fluid-like forces during substrate interaction depending on the magnitude of normal and shearing forces~\underline{\cite{neddermanbook}}.

Granular material~\underline{\cite{jaeAnag}}, which is common in nature, proves a good model substrate for studying locomotion on deformable ground~\underline{\cite{cheAumb09,malAdin09,cheAumb10,mazAgra10}}. Granular media display the solid-fluid transition during limb-ground interaction, and are at the same time simpler than most other deformable and/or flowing substrates. Another advantage of using granular media as test substrates is that their mechanical properties are readily and precisely controlled by varying the volume (packing) fraction of their constituent particles (see Experiments section).

There is increasing evidence that small, legged machines can have greater maneuverability than large wheeled vehicles in many natural environments~\underline{\cite{raibert86}}. For example, recent studies of the legged RHex-class robot SandBot~\underline{\cite{cheAumb09,cheAumb10}} demonstrate that effective locomotion on deformable substrates like granular media is possible when stride frequencies and limb kinematics are appropriate for ground properties. In particular, these studies suggest that the ability of limbs to generate adequate ground thrust relative to gravitational and inertial forces is the key to effective locomotion on deformable substrates. Other factors, such as limb morphology and compliance~\underline{\cite{moritz2003human}} and neurosensory control,~\underline{\cite{daley2006running}} may also substantially effect locomotion on deformable substrates.

The DASH and RoACH robots used in this study are small, biologically inspired, six-legged, lightweight ($\sim10$~cm, $\sim20$~g) robots. Because their mass is $\sim100$ times smaller than earlier top performing devices like the RHex class of robots~\underline{\cite{saranli2001rhex}}, DASH and RoACH should traverse varied terrains with higher mobility and lower energy cost than any previous legged devices due to their potentially much larger thrust-to-gravity ratio. Although similar in size and composition, each robot has distinct limb morphology and kinematics and, as such, provide excellent test platforms for assessing benefits and tradeoffs of these factors on a variety of substrates. In addition, the robots are simple and inexpensive to construct, have relatively high power densities, and can survive collisions with the ground at their terminal velocities~\underline{\cite{DASH}}.

To probe the locomotor capabilities of small lightweight robotic platforms on differing terrains and to examine benefits and tradeoffs of different limb morphologies and kinematics, we systematically test the locomotor performance (average speed, mean mechanical power and cost of transport, and stability) of DASH and RoACH as they move on two distinct types of substrates, hard ground and deformable granular media. We vary by open loop control the robot stride frequency on prepared substrates of well defined mechanical properties and compare performance across robotic platforms and substrates.

\section{Robotic Platforms}

The robotic platforms featured in these experiments are small, lightweight ($\sim10$~cm, $\sim20$~g) hexapedal robots that utilize alternating tripod gaits for horizontal locomotion. They both feature drive mechanisms which use a single $6$~mm brushed DC motor (I-Fly Vamp) to drive all six limbs. Onboard lithium 90~mAh polymer battery (Fullriver), Bluetooth communications (Roving Networks RN-41), microcontroller (Microchip PIC dsPIC33), and motor driver allow the robots to run preprogrammed commands autonomously while sensors detect back electromotive force (back EMF) in the DC motor.

\subsection{Fabrication}

\label{sec:scm}
The robot fabrication process closely follows the smart composite microstructures (SCM) process\underline{\cite{Wood08.1}}, but instead of pre-impregnated composite fiber laminate, we substitute lightweight, double sided coated posterboard (Nature Saver posterboard, Officemax.com). For the polyimide flexure layer, we substitute a thicker polyethylene terephalate (PET) film because thermal compatibility of the film is less of a concern than in the SCM process. In addition, the PET is less inert so it bonds better to various adhesive polymers. Just as in the SCM process, the posterboard is cut using a precision laser (VersaLaser VL$200$) according to a $2$-dimensional design drawing. However, unlike with carbon fiber composites, the posterboard can be easily cut with an infrared laser. The steps for the prototyping process are outlined below with the step numbers corresponding to Fig.\ \hyperlink{fig1}{\underline{1}}:

\Needspace{1.6in}\begin{enumerate}
	\item The process begins with a sheet of posterboard.
	\item Gaps are laser-cut into the posterboard where flexures will be located in the final structure.
	\item Two layers of adhesive polymer with one layer of flexural polymer (high melting point) are placed between two sheets of posterboard.
	\item The resulting structure is rolled through heated rollers to apply pressure and melt the adhesive layers, bonding the structure.
	\item Outlines of parts are then laser cut, releasing the parts from the original sheet.
\end{enumerate}

\manuscriptfigure{1}{width=\linewidth,height=4.6in,keepaspectratio}{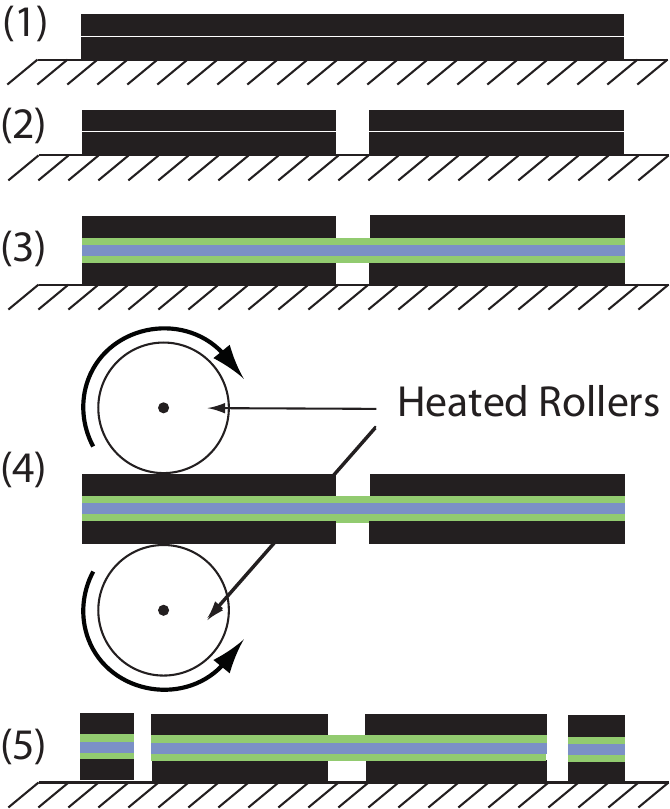}{Step-by-step illustration of the posterboard link and hinge fabrication process\underline{\cite{Hoover08}}.}

Because the posterboard, unlike pre-impregnated composites, contains no adhesive, a simple thermoset polymer film for the flexure layer is insufficient. We must either add adhesive layers separately, or use a polymer film with an integrated adhesive to bond the structure. A carrier film with integrated adhesive is convenient because it is essentially monolithic. However, the standard thicknesses of such films limit the ability of the designer to change the stiffness of flexure hinges by using different thickness films. While it is possible to use more than one layer of such a film, the interaction of the adhesive layers between the carrier film layers creates a noticeable viscoelastic response from the flexure hinges. In order to retain flexibility for the designer in the choice of flexure thickness, we have chosen to add a separate adhesive between the posterboard and the flexure film. For the the flexure film we use $25$~$\mu$m PET depending on the desired flexure stiffness. For the adhesive layer we use Octiva hot mount adhesive (gbcconnect.com), an unsupported, heat-activated mounting adhesive. Before the posterboard robot parts are cut, the hot mount is applied to one face of the posterboard and the liner film detached.

This fabrication method is appropriate for fabricating entire functional robots or simply mocking up a small mechanism to be integrated into a larger robot design. The process makes it possible to construct very lightweight robots with complex spatial kinematics that exhibit relatively high power densities. The posterboard SCM process also enables the designer to quickly try out a folded design and explore features such as alignment tabs or self-jamming flexures that can aid significantly in the final assembly process. For example, the designs presented below can be cut out and assembled in a matter of two to three hours. These hexapod crawler designs demonstrate the complicated 3D mechanisms that can result from folding a $2$D design using this approach.

\subsection{RoACH}
\label{sec:roach}

The Robotic Autonomous Crawling Hexapod (RoACH, Fig.\ \hyperlink{fig2}{\underline{2}}) is a six-legged robot fabricated using the process described above. The kinematic design of the robot uses a combination of slider crank and four-bar linkages to create an alternating tripod gait that can be actuated using a single rotary input connected to a crank. The abduction and adduction degree of freedom is actuated by a slider crank mechanism. Protraction and retraction of the leg are achieved using a parallel four-bar linkage at the hip. Diagrams of the kinematics hip linkages of RoACH are shown in Fig.\ \hyperlink{fig3}{\underline{3}}. As configured for this experiment, RoACH had a mass of $24.9$~g.

\manuscriptfigure{2}{width=\linewidth,height=4.6in,keepaspectratio}{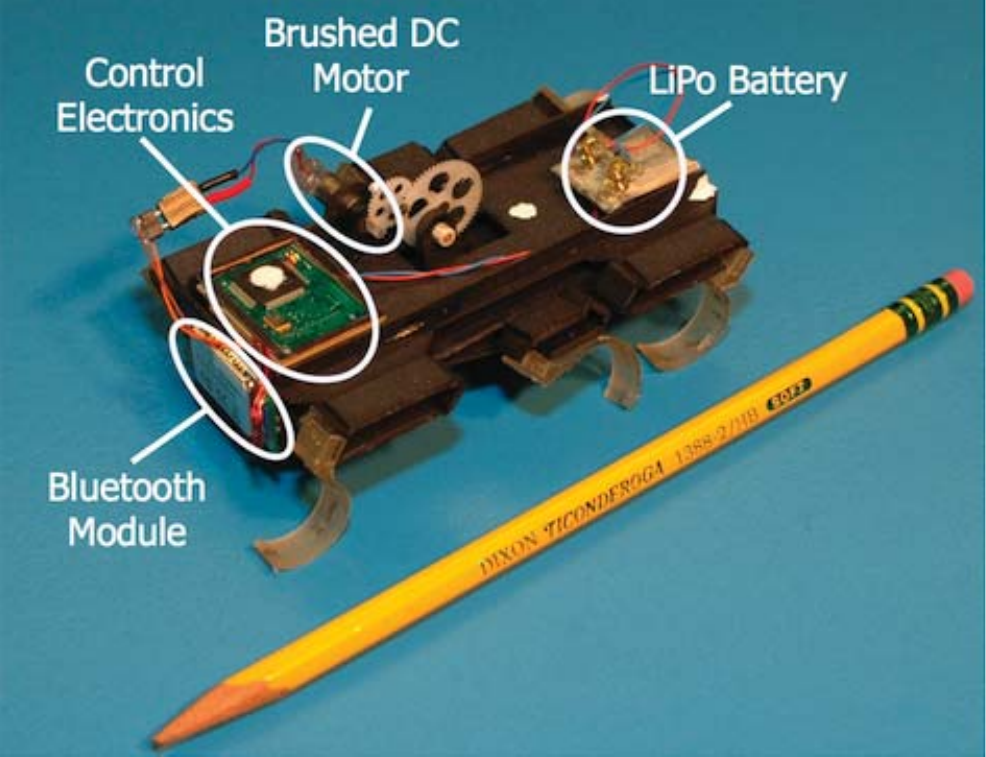}{The RoACH robot shown with component parts of onboard power and electronics highlighted.}

\manuscriptfigure{3}{width=\linewidth,height=4.6in,keepaspectratio}{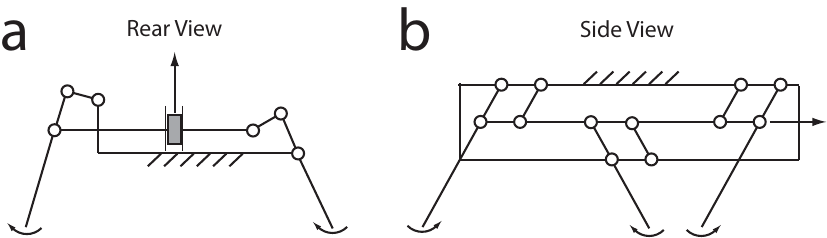}{Linkage kinematics for RoACH's alternating tripod gait. (a) Vertical translation of the center link causes one tripod to engage the ground while the other is lifted. (b) Fore-aft translation of the central link causes the front and rear legs to rotate forward and the middle leg to rotate backward. Driving the center link in a circle in the sagittal plane creates the alternating tripod gait.
}

\subsection{DASH}
\label{sec:dash}

A second robotic platform developed using the posterboard SCM process is the Dynamic Autonomous Sprawled Hexapod (DASH, Fig.\ \hyperlink{fig4}{\underline{4}})~\underline{\cite{DASH}}. Like RoACH, DASH employs an alternating tripod gait which is driven by a single rotary input. The structure of DASH creates an oar-like transmission in which the circular input in the sagittal plane creates circular leg trajectories. The fore-aft displacement of the motor output rotates the four-bar hip mechanisms, protracting one tripod and retracting the other. The vertical motion of the motor output angles the orientation of the hip four-bar mechanism which simultaneously lifts one tripod and lowers the other. Diagrams of the kinematics hip linkages of DASH are shown in Fig.\ \hyperlink{fig5}{\underline{5}}. These vertical and fore-aft degrees of freedom are coupled together to enable a circular leg trajectory. With the battery and all onboard electronics for this experiment, DASH had a mass of $22.1$~g.

\manuscriptfigure{4}{width=\linewidth,height=3in,keepaspectratio}{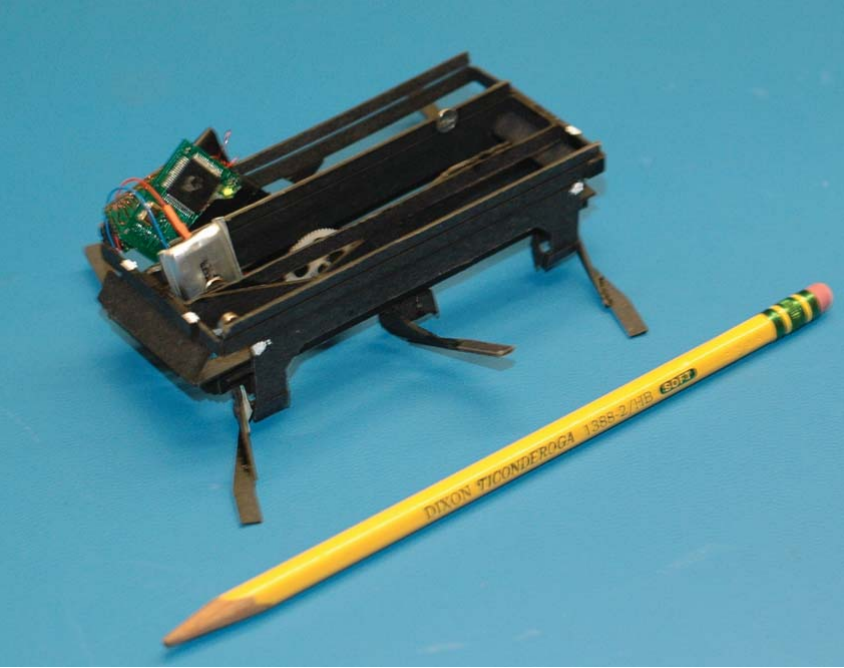}{The DASH robot with component parts of onboard power and electronics (same as RoACH).}

\manuscriptfigure{5}{width=\linewidth,height=4.6in,keepaspectratio}{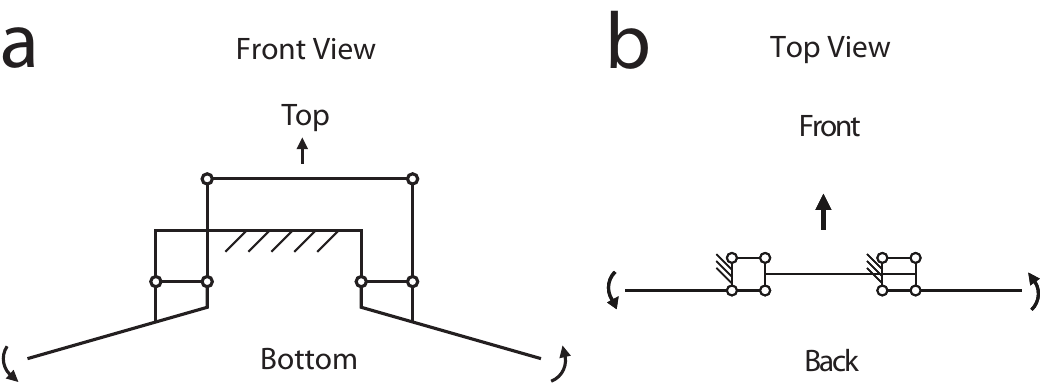}{Linkage kinematics of DASH. (a) Vertical translation of top link causes one tripod to lift and the other to lower. (b) Fore-aft motion of the center link causes one tripod to rotate forward and the other to rotate backward.
}

While DASH and RoACH share a number of features such as their manufacturing process, electronics, motor, and even an alternating tripod gait, they have significant morphological differences. The most notable among these differences is their posture. DASH has a sprawled posture, with its legs directed outward nearly horizontally; RoACH has an upright posture with its legs directed almost vertically downward.

\section{Experiments}

\subsection{Experiment Design}

The volume (packing) fraction $\phi$ of a granular material is defined as the ratio of the solid volume of the grains forming the media to the volume they occupy. Higher volume fraction indicates grains are more densely packed. In nature and under gravity, dry, spherical, granular media exist in a range of volume fraction $0.55 < \phi < 0.64$ ~\underline{\cite{dickinson1994low}}. Granular media of different volume fraction often exhibit very different mechanical properties. For example, we find that as a bed of poppy seeds is compacted from $\phi = 0.58$ to $\phi = 0.64$, reaction force during constant velocity vertical penetration increases approximately five-fold. The dynamics of ballistic impact under gravity also differ dramatically~\underline{\cite{umbAgol10}} with varying volume fraction.

An apparatus known as a fluidized bed can precisely control the volume fraction and thus the mechanical properties of granular media and has been used successfully for this purpose in physics and locomotion biology studies~\underline{\cite{golAswi06,goldmanKorff06,cheAumb09}}. A sufficient counter flow of air through the grain bed against gravity brings the grains into a fluidized state with neutral buoyancy. Upon cessation of the air flow, the granular media settles to a loosely packed state of low volume fraction. Subsequent repeated air pulses (or vibration aided by small air flow) compacts the grains to higher $\phi$.

In this study, we use a fluidized bed trackway to prepare the granular media in a loosely packed state ($\phi = 0.580$). As previously discovered~\underline{\cite{cheAumb09}}, locomotion at low $\phi$ is more challenging than at high $\phi$ as limbs can easily fluidize the grains during ground interaction. By comparing the performance of small lightweight robotic platforms on rigid ground and on loosely packed granular media we sample two representative extremes from the spectrum of terrestrial substrates a mobile device could encounter. By comparing performance across RoACH and DASH platforms, we assess the benefits and tradeoffs of different limb morphology and kinematics on these two distinct substrates.

\subsection{Experimental Setup}

Figure \hyperlink{fig6}{\underline{6a}} illustrates the experimental setup. Hard ground was realized with a piece of drywall ($120$~cm $\times50$~cm $\times1$~cm) with uniform surface properties. Loosely packed granular media was realized using poppy seeds prepared with $\phi = 0.580$ in our $2.5$~m long, $0.5$ m wide fluidized bed trackway. Poppy seeds were chosen as the granular material because their size is close to natural sand \underline{\cite{bagnold}} and their density ($\approx1.0$ kg/cm$^3$) low enough to be fluidized. The fluidized bed trackway has a horizontal porous plastic (Porex) flow distributor (thickness $0.64$~cm, average pore size $90$~$\mu$m) on which the poppy seeds rest. By blowing air upward through the distributor with four $300$ LPM leaf blowers (Toro) at maximum power, the entire bulk of granular media was fluidized into the bubbling regime. As the flow was slowly reduced to zero, the poppy seeds came to rest in a loosely packed state. The air flow across the fluidized bed is measured with an anemometer (Omega Engineering FMA-$900$-V) and is uniform to within $10$ percent. Volume fraction was calculated as $\phi$ = $m$/$\rho$$V,$ where $m$ is the total mass of the seeds, $\rho$ the average solid density of each grain and $V$ the bed volume.

\manuscriptfigure{6}{width=\linewidth,height=4.6in,keepaspectratio}{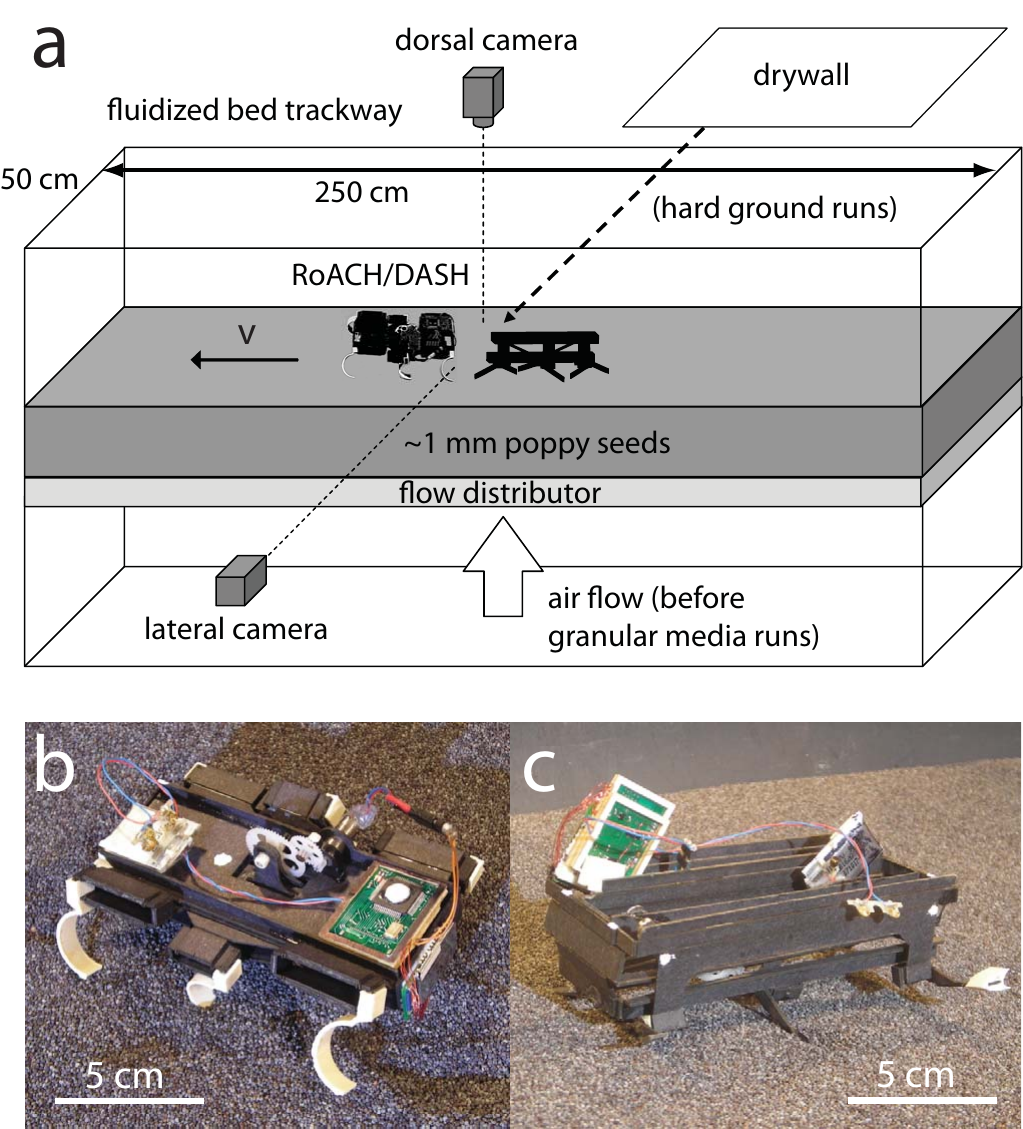}{(a) Illustration of the experimental setup (not to scale). A fluidized bed trackway prepares the granular material, $\sim1$~mm diameter poppy seeds, into loosely packed states to create a deformable substrate with well defined mechanical properties. Hard ground is simulated by a piece of drywall. Dorsal and laterals cameras record high speed videos of the kinematics as DASH/RoACH moves on the substrate. Robot sensors record the back EMF and the control reference signal. (b) RoACH and (c) DASH stands on loosely packed poppy seeds.
}

The substrate was prepared before each run: for hard ground runs, the drywall rested horizontally on the flat granular media surface, and was wiped with a brush to remove any contaminants from the previous run (poppy seeds occasionally got onto the drywall when the robots ran off the end); for granular media runs, the poppy seeds were prepared into a loosely packed state with volume fraction $\phi = 0.580$ using the fluidized bed trackway (air flow is turned off before the robot is run). The RoACH/DASH (Fig.~\hyperlink{fig6}{\underline{6b,c}}) robot was carefully placed on the substrate at the same end of the trackway with one tripod entirely in contact with the surface while the other remained lifted above the substrate. On the granular media, DASH was placed with its stance with all six legs touching the surface. RoACH is run with the curvature of the legs pointing forward (rather than backward as is the case with the RHex and Sandbot robots) because this was observed to improve performance on hard ground.

A Python graphic user interface (GUI) is used to configure the robot via a $230$~Kbps Bluetooth communication link. The robot's DC motor is controlled using a pulse width modulated (PWM) voltage, and the control reference signal sent to the robot is strictly an open loop, feedforward specification of the duty cycle of that signal (under open loop stride low frequency $f < \sim5$ Hz is not possible). The PIC microcontroller includes a 10-bit analog to digital converter that is used to sample the back EMF of the motor when the motor is switched off during the idle phase of the PWM cycle.  Those measurements together with the voltage control reference signal are streamed back to the computer base station running the GUI for plotting and saving.

Simultaneous dorsal and lateral videos of the robot motion are recorded at 1280$\times$1024 resolution and 250~fps with high speed video cameras (AOS Switzerland) to capture the kinematics of the robot and its interaction with the substrate. The back and side of each robot is marked with reflective material (WiteOut). Lighting is provided by four overhead work lights, a side work light (HomeDepot), and a high intensity HMI light (Arri). Videos and robot sensor data are synchronized by identifying the beginning of robot motion in the video and the beginning of the recording of the robot sensor data (which corresponds to the start of robot motion).

The stride frequency $f$ of each run is measured from the lateral video frames. The average speed $\overline{v}$ of each run is calculated using distance traveled and elapsed time from the dorsal video frames. The mean mechanical power $\overline{P}$ is averaged from $P = N\omega$, where $N = K_tI$ is the motor torque, and $\omega = 2\pi V_{EMF}/K_e$ is the motor angular velocity, with the motor torque constant $K_t = 0.00683$~Nm/A, and the motor back EMF constant $K_e = 0.15$~V/Hz. The motor current $I = (V_{ref}-V_{EMF})/R$ is determined from the back EMF $V_{EMF}$ and the voltage control reference signal $V_{ref}$ recorded by the motor sensors using the motor resistance $R = 4.2$~$\Omega$. The mean mechanical cost of transport is given by $\overline{COT} = \overline{P}/m_R\overline{v}$, where $m_R$ is the mass of the robot. Both the mechanical power and cost of transport characterize only the mechanical energy and do not include heat dissipation in the electronics and the motor. We verified that the calculated motor frequency $\omega/2\pi$ and the stride frequency $f$ determined from the lateral video are the same (within $5\%$ deviation). Averaging of all data was performed on the second half of each run where the robot reached a steady state speed.

\section{Results and Discussions}

Figure \hyperlink{fig7}{\underline{7}} shows average speed as a function of stride frequency for both robotic platforms on both substrates. On hard ground, both DASH and RoACH achieve high speeds, e.g.\ at $f \approx 15$~Hz, $\overline{v} \approx 80-100$~cm/s or $\sim10$~bodylength/s (as compared to the peak performance of $\sim1$~bodylength/s for the RHex class robots). DASH displays a roughly proportional $\overline{v}$-$f$ relationship (solid green line is a linear fit with zero intercept) indicative of a constant stride length for all stride frequencies. On the other hand, RoACH has a more irregular $\overline{v}$-$f$ pattern on hard ground, which may be a result of RoACH's more erratic vertical bouncing dynamics as observed in the lateral videos.

\manuscriptfigure{7}{width=\linewidth,height=4in,keepaspectratio}{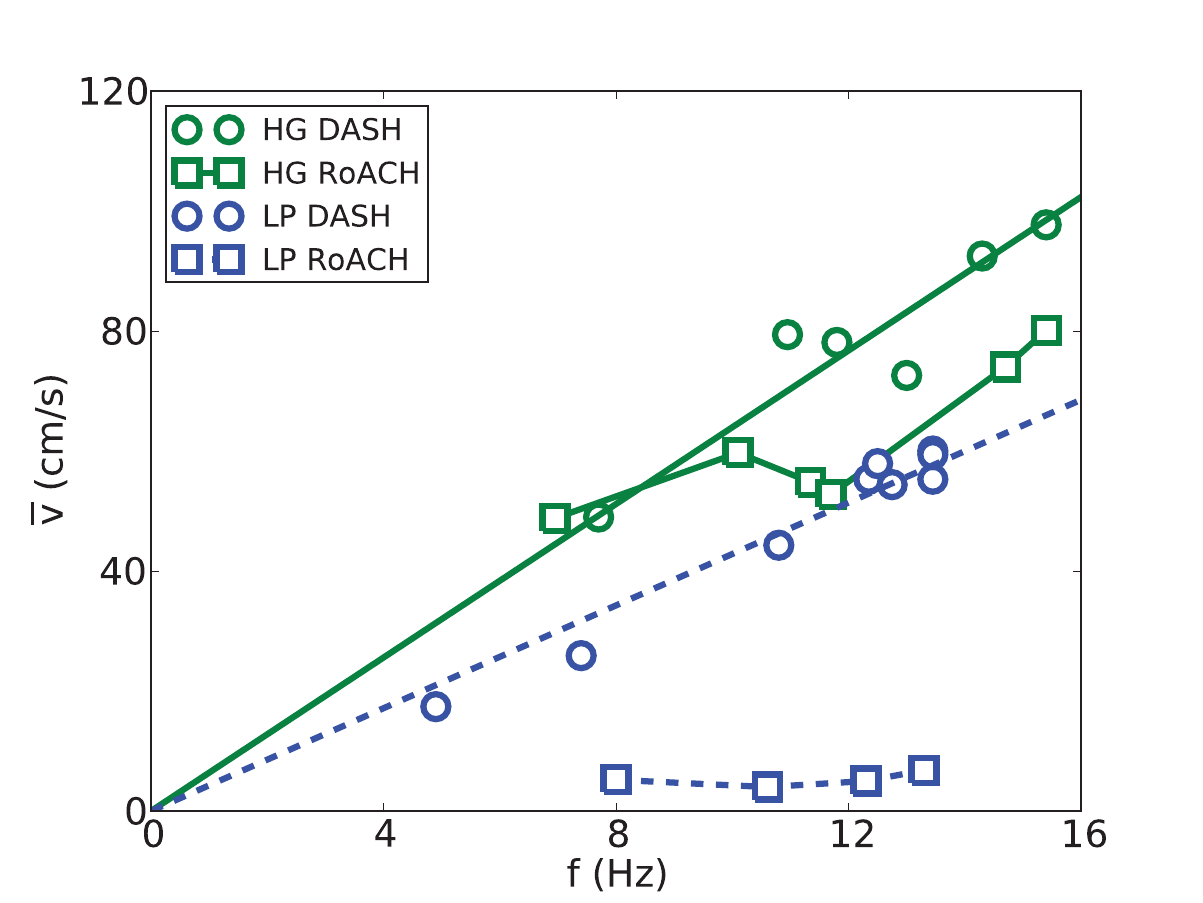}{Average speed as a function of stride frequency of DASH (circles) and ROACH (square) on hard ground (HG, green) and loosely packed granular media (LP, blue). Lines are linear fits to the DASH data assuming a linear $\overline{v}$-$f$ relationship with an $f$ intercept of zero.
}

On loosely packed granular media, both robots suffer substantial performance loss. DASH's speed drops by $\sim40\%$ for all frequencies, but maintains a proportional $\overline{v}$-$f$ relationship (dashed blue line). The slope of the $\overline{v}$-$f$ relationship is lower as compared to hard ground data, suggesting that there is more slip during each stride, although stride length remains independent of stride frequency. On the other hand, RoACH suffers near total ($> 90\%$) performance loss, with $\overline{v} \approx 5$~cm/s and roughly independent of frequency.

The difference in performance loss can be explained by the different limb morphology and kinematics of both robots. Closer observation of the robot kinematics from lateral high speed videos show that RoACH, due to its upright posture combined with its compliant limbs and kinematics (the curvature of the legs pointing forward), cannot effectively retract its limbs from the grains; instead, RoACH's limbs penetrate deep into the substrate and constantly move back and forth within the fluidized grains with the body lying on the surface, resulting in tripod interference and large belly drag. The inability to retract limbs causes RoACH to ``swim'' at low speeds in fluidized granular media via frictional drag on the limbs similar to how SandBot swims in granular media~\underline{\cite{cheAumb09}}. DASH's sprawled limb posture and kinematics, on the other hand, prevents it from getting trapped in the swimming regime for the parameters studied here and enables it to achieve $\sim5$~bodylength/s on loosely packed poppy seeds, a huge improvement from SandBot's $\sim0.1$ bodylength/s~\underline{\cite{cheAumb09}}. This confirmed our speculation that on deformable substrates such as granular media, small lightweight robots of DASH's scale can achieve higher mobility because of their high thrust-to-gravity ratio provided that limb morphology and kinematics are properly designed.

It is also worth noting from Fig.\ \hyperlink{fig7}{\underline{7}} that for DASH, run-to-run deviation from the linear $\overline{v}$-$f$ relationship is more pronounced on hard ground than on loosely packed granular media. The irregular $\overline{v}$-$f$ relationship for RoACH on hard ground also reduces to a roughly constant relationship on loosely packed granular media. We speculate that such reduction in run-to-run variation in speed originates from differences in stability on the two distinct substrates. On hard ground, while mobility is higher for both robots because of the larger ground reaction forces and reduced drag, their rapid bouncing motion can suffer greatly from yaw, pitch, and roll instabilities caused by non-periodic impacts and the large kinetic energy associated with bouncing dynamics, which may result in large run-to-run variation in speed. On granular media, instability is rare because the ``softer'' dissipative granular media serves as a ``buffer'' to absorb any abrupt forces and accommodate a smoother motion at the cost of reduction in speed. This is consistent with previous observations in SandBot~\underline{\cite{cheAumb09}}.

\manuscriptfigure{8}{width=\linewidth,height=4.6in,keepaspectratio}{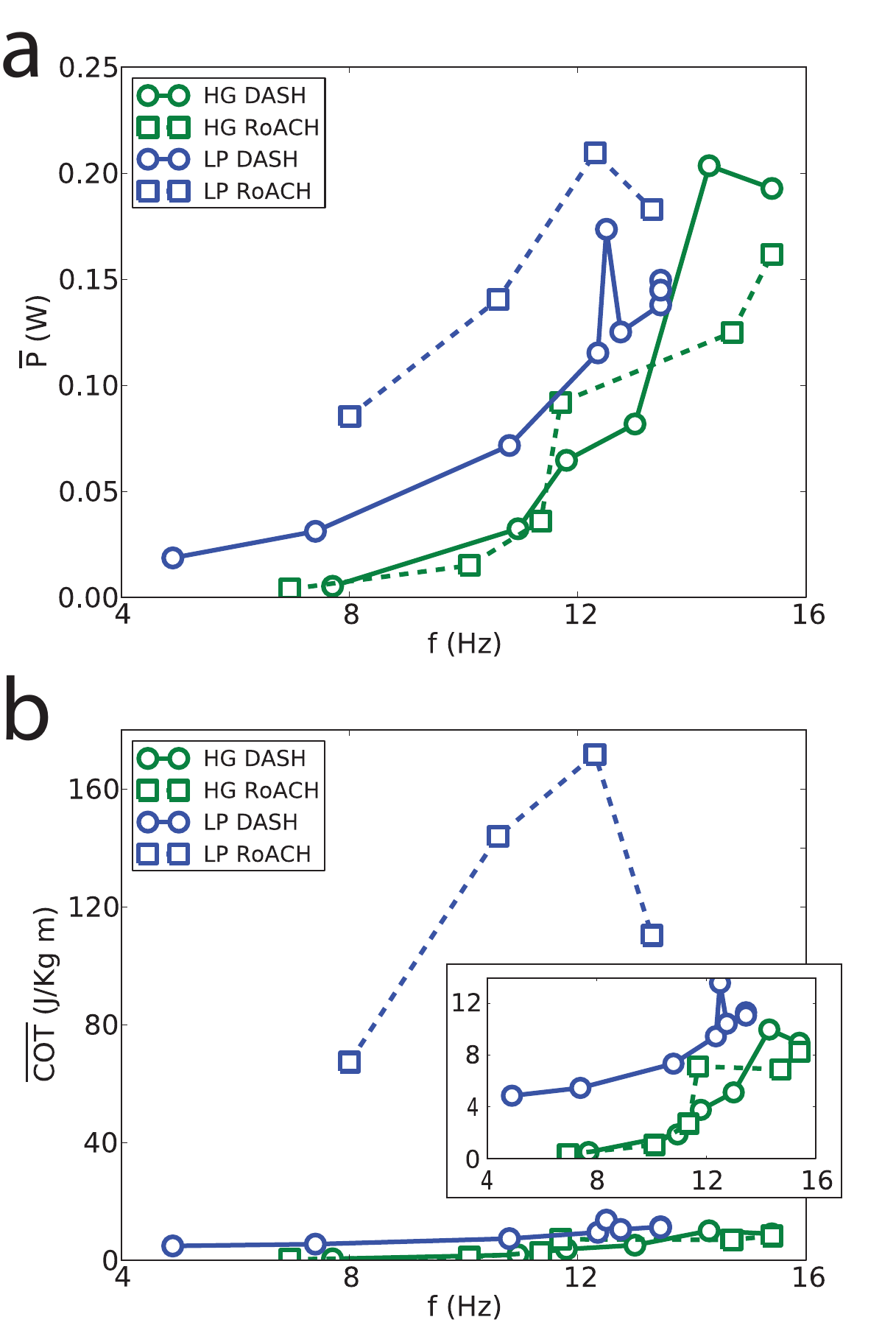}{Power consumption characteristics of DASH and RoACH moving on hard ground (green) vs. loosely packed granular media (blue). (a) Mean mechanical power $\overline{P}$ as a function of stride frequency. (b) Mean mechanical cost of transport $\overline{COT}$ as a function of frequency. Inset highlights the data in the main figure which is obscured by high $\overline{COT}$ of RoACH on loosely packed granular media.
}

The distinct differences between foot-ground interaction with hard ground and with soft granular media is further revealed by the power consumption characteristics. As shown in Fig.\ \hyperlink{fig8}{\underline{8a}}, the mean mechanical power $\overline{P}$ for both robots is consistently higher on loosely packed granular media than on hard ground at all frequencies, with a smaller increase for DASH than for RoACH. This supports our speculation that DASH's sprawled limb posture and kinematics is more effective for locomotion on granular media. Also noteworthy is the faster than linear increase of $\overline{P}$ with $f$ for both robots on both substrates. This may be an indication that the ground reaction force increases with frequency. Since in granular media, force is roughly rate-independent~\underline{\cite{wieghardt75,malAdin09}}, this implies that the granular reaction force may be dominated by the inertial force of translating the robot forward at each step (similar to~\underline{\cite{mazAgra10}}). In such locomotor mode we expect the power to increase as $f^3$ and the data is not inconsistent with this prediction. It is noteworthy that even on loosely packed granular media, the mechanical power of DASH and RoACH is on the order of $10^{-1}$ W, on the order of $100$ times smaller than RHex's $10^1$ W on hard ground~\underline{\cite{saranli2001rhex}}.

Fig. \hyperlink{fig8}{\underline{8b}} shows the mean mechanical cost of transport of both robots on both substrates. The most striking feature of the data is that RoACH on loosely packed granular media has very large mechanical cost of transport as compared to other cases ($\sim10$ times larger). This results directly from RoACH's very slow ``swimming'' motion on loosely packed granular media, as $COT \sim v^{-1}$. During swimming motion, limbs constantly translate within fluidized grains and only generate net forward motion when the frictional drag on the limbs overcome that on the body. Most of the mechanical work is dissipated by the fluidized granular material which acts like a highly dissipative frictional fluid in which force is rate-independent~\underline{\cite{malAdin09}} (this is different from the rate-dependent viscous dissipation of Newtonian fluids at low Reynolds number). The inset in Fig. \hyperlink{fig8}{\underline{8b}} shows that $\overline{COT}$ increases with $f$ faster than linearly for all other cases. Comparing DASH and RoACH, we see that locomotion speed and energy consumption on granular media are both greatly improved when the locomotor avoids operating in the swimming regime and instead makes use of solidified granular media by adjusting morphology, kinematics, frequency, etc. This confirms the importance of utilizing the solidification properties of granular media for effective locomotion discovered in previous studies~\underline{\cite{cheAumb09,cheAumb10,mazAgra10}}.

\section{Conclusions}
\label{sec:conclusions}

To probe the capability and assess the benefits and tradeoffs of various limb morphology and kinematics of small legged robotic platforms on a variety of terrains, we systematically tested the performance of the DASH and RoACH robots with different limb morphology and kinematics on hard ground and loosely packed granular media. We investigated how speed and power consumption characteristics vary with stride frequency under open loop control and compared performance across robotic platforms and substrates. We found that both robots suffer performance loss in the form of reduction in speed and increase in mechanical power and cost of transport as substrates weaken from rigid hard ground to deformable granular media, while the instability intrinsic to bouncing motion on hard ground is largely removed by the dissipative nature of granular media. We also found that differences in robot limb morphology and kinematics have a large impact on the platforms susceptibility to performance loss and instability. Future work could test sensitivity of performance to changes in limb morphology and kinematics by systematically varying these factors in analogy with the variation of SandBot's kinematics~\underline{\cite{cheAumb10}}.

Our study is among the first to investigate how the performance of small lightweight legged robots differs as the substrate changes from rigid to deformable ground. Our results show that when moving on deformable ground such as granular media, even for lightweight platforms with relatively large thrust to gravity ratio, precaution needs to be taken to avoid limbs getting caught and being trapped in the swimming regime. Traversing hard ground, on the other hand, requires careful management of kinetic energy and ground reaction dynamics for legged devices to minimize yaw, pitch, and roll instabilities.

Careful study of small physical models such as DASH and RoACH can also inform locomotion biology in understanding how small animals appear to move effortlessly across a diversity of complex substrates~\underline{\cite{golAche06,spaAgol}}. Further studies of this kind can lead to systematic empirical laws (and ultimately comprehensive theory) of limb-ground interaction with a spectrum of substrates ranging from rigid ground to deformable grounds like granular media. Such advanced understanding will inform limb design and control methodologies, similar to the well-developed models of aerial and aquatic craft. Future devices will begin to perform at levels comparable to organisms; these capabilities could be used for more efficient and capable exploration of challenging terrestrial environments ({\em e.g.\ } battlefields, building debris, and natural disasters sites).

\section*{Acknowledgments}     

We thank Sam Burden for help with data analysis. A.M.H. thanks Xiao-Yu Fu for his tireless hard work in fabricating leg designs and building robots in the days leading up to the experiments. This work was supported by the Burroughs Wellcome Fund (D.I.G., C.L., and P.B.U.), the Army Research Laboratory (ARL) Micro Autonomous Systems and Technology (MAST) Collaborative Technology Alliance (CTA) under cooperative agreement number W911NF-08-2-0004 (D.I.G., P.B.U., R.S.F., and A.M.H.), as well as the National Science Foundation Grant EEC-0832819 (P.B.).

\section*{Supporting movies}
{\fontsize{10}{13}\selectfont\begin{enumerate}\setlength{\itemsep}{3pt}
\item \href{https://www.youtube.com/watch?v=e7Ge2H_iTqk&ab_channel=TerradynamicsLab}{\underline{Video 1: DASH on solid ground}}
\item \href{https://www.youtube.com/watch?v=6Z8dij-qVQU&ab_channel=TerradynamicsLab}{\underline{Video 2: RoACH on solid ground}}
\item \href{https://www.youtube.com/watch?v=llh3iS_2_Kg&ab_channel=TerradynamicsLab}{\underline{Video 3: DASH on granular media}}
\item \href{https://www.youtube.com/watch?v=zaXXAml-jO4&ab_channel=TerradynamicsLab}{\underline{Video 4: RoACH on granular media}}
\item \href{https://www.youtube.com/watch?v=kvHC_-FUUQA&ab_channel=TerradynamicsLab}{\underline{RoACH running on granular media with C-legs}}
\item \href{https://www.youtube.com/watch?v=3vGz8ocQ6as&ab_channel=TerradynamicsLab}{\underline{RoACH stuck on granular media with reversed C-legs}}
\end{enumerate}}
\end{document}